\documentclass[useAMS,usenatbib]{mn2e}
\usepackage{graphicx}
 
\def\be{\begin{equation}} 
\def\ee{\end{equation}}

\def\HI{\hbox{H~$\scriptstyle\rm I\ $}} 
\def\HII{\hbox{H~$\scriptstyle\rm II\ $}}

\def\gsim{\lower.5ex\hbox{\gtsima}} 
\def\lsim{\lower.5ex\hbox{\ltsima}} \def\gtsima{$\; \buildrel > \over 
\sim \;$} \def\ltsima{$\; \buildrel < \over \sim \;$} \def\prosima{$\; 
\buildrel \propto \over \sim \;$} \def\gsim{\lower.5ex\hbox{\gtsima}} 
\def\lsim{\lower.5ex\hbox{\ltsima}} 
\def\simgt{\lower.5ex\hbox{\gtsima}} 
\def\simlt{\lower.5ex\hbox{\ltsima}} 
\def\simpr{\lower.5ex\hbox{\prosima}} \def\la{\lsim} \def\ga{\gsim}

\def\gtsima{$\; \buildrel > \over \sim \;$} 
\def\ltsima{$\; \buildrel < \over \sim \;$} 
\def\gsim{\lower.5ex\hbox{\gtsima}} 
\def\lsim{\lower.5ex\hbox{\ltsima}} 
\def\simgt{\lower.5ex\hbox{\gtsima}} 
\def\simlt{\lower.5ex\hbox{\ltsima}} 
\def\simpr{\lower.5ex\hbox{\prosima}} 
\def\la{\lsim} 
\def\ga{\gsim}

\def\E3{{\cal E}_{\rm g}^{III}}

\title[LAE evolution]{Lyman Alpha Emitter Evolution in the Reionization Epoch } 
\author[P. Dayal et al.]{P. Dayal$^{1}$\thanks{E-mail: 
dayal@sissa.it (PD)}, A. Ferrara$^{2}$, A. Saro$^{3,4}$, R. Salvaterra$^{5}$, S. Borgani$^{3,4,6}$ \& L.Tornatore$^{3,4,6}$  \\ 
$^{{1}}$ SISSA/International School for Advanced Studies, Via Beirut 2-4 Trieste, Italy, 34014\\ 
$^{2}$ Scuola Normale Superiore, Piazza dei Cavalieri 7, 56126 Pisa, Italy \\
$^{3}$ Dipartimento di Astronomia dell'Universita di Trieste, via Tiepolo 11, 34131 Trieste, Italy\\
$^{4}$ INFN, National Institute for Nuclear Physics, Trieste, Italy \\
$^{5}$ INAF/Osservatorio Astronomico di Brera, via Emilio Bianchi 46, 23807, Merate (LC), Italy\\
$^{6}$INAF/Osservatorio Astronomico di Trieste, via Tiepolo 11, 34131 Trieste, Italy}

\begin{document} 
 
\date{Received 2009 March 1; in original form 2009 March 1} 
 
\pagerange{\pageref{firstpage}--\pageref{lastpage}} \pubyear{2002} 
 
\maketitle 
 
\label{firstpage} 
 
\begin{abstract} 
Combining cosmological SPH simulations with a previously developed Ly$\alpha$ production/transmission model and the Early Reionization Model (ERM, reionization ends at redshift $z \sim 7$), we obtain Ly$\alpha$ and UV Luminosity Functions (LFs) for Lyman Alpha Emitters (LAEs) at $5.7 \le z  \le 7.6$. Matching model results to observations at $z\sim 5.7$ requires escape fractions of Ly$\alpha$, $f_\alpha=0.3$, and UV (non-ionizing) continuum photons, $f_c=0.22$, corresponding to a color excess, $E(B-V)=0.15$. We find that (i) $f_c$ increases towards higher redshifts, due the decreasing mean dust content of galaxies, (ii) the evolution of $f_\alpha/f_c$ hints at the dust content of the ISM becoming progressively inhomogeneous/clumped with decreasing redshift. Using the model assumptions, clustering of sources has little effect on the Ly$\alpha$ LF for a cosmic hydrogen neutral fraction $\chi_{HI} \le 10^{-4}$, a value attained at $z \leq 6.6$ in the ERM. However, during the initial reionization phases ($z \simgt 7$) the clustering photoionization boost becomes important. We quantify the physical properties of observed LAEs and their redshift  evolution, for which we give handy analytical fitting functions. Halo (stellar) masses are in the range $10.0 < \log M_h < 11.8$ ($8.1 < \log M_* < 10.4$) with $M_h \propto M_*^{0.64}$. The star formation rates are $\dot M_* =3-120\, {\rm M_\odot \, yr^{-1}}$, mass-weighted mean ages are $t_* > 20$ Myr at all redshifts, while the mean stellar metallicity increases from $Z=0.12\,{\rm Z_\odot}$  at $z \sim 7.6$ to $Z=0.22\,{\rm Z_\odot}$ at $z \sim 5.7$; both $t_*$ and $Z$ positively correlate with stellar mass. The brightest LAEs are all characterized by large $\dot M_*$ and intermediate ages ($\approx 200$ Myr), while objects in the faint end of the Ly$\alpha$ LF show large age and star formation rate spreads. With no more free parameters, the Spectral Energy Distributions of three LAE at $z\sim 5.7$ observed by Lai et al. (2007) are well reproduced by an intermediate age ($182-220$ Myr) stellar population and the above $E(B-V)$ value. The model uncertainties, mostly related to the simplified treatment of dust and to the possible effects related to gas outflow/infall, are discussed along with their impact on the results. 
\end{abstract} 
 
\begin{keywords}
 methods:numerical - galaxies:high redshift - luminosity function - intergalactic medium - cosmology:theory 
\end{keywords} 
 
\section{Introduction}
\label{intro}
Over the past few years, Lyman Alpha Emitters (LAEs) have rapidly been gaining popularity as probes of cosmic reionization for two primary reasons. Firstly, specific signatures like the strength, width of the Ly$\alpha$ line (1216 \AA) and the continuum break bluewards of it make the detection of LAEs unambiguous. Secondly, since Ly$\alpha$ photons are highly sensitive to the presence of neutral hydrogen, their attenuation can be used to put constraints on the ionization state of the IGM.  

Since these are also amongst the earliest galaxies to have formed, they represent superb probes of the properties and evolution of early galaxy populations. This is important since mechanical, chemical and radiative feedback from these galaxies will determine the properties of galaxies formed later on. Obtaining the star formation rates (SFR), metallicity, initial mass function (IMF) of these galaxies are hence imperative in understanding galaxy evolution.

Advances in instrument sensitivity have enabled observers to push the observable frontier to increasingly high redshifts. There are now hundreds of confirmed LAEs at $z\sim2. 25$ (Nilsson et al. 2008), $z\sim3$ (Cowie \& Hu 1998; Steidel et al. 2000; Matsuda et al. 2005; Venemans et al. 2007; Ouchi et al. 2008), $z\sim4.5$ (Finkelstein et al. 2007), $z\sim5.7$ (Malhotra et al. 2005; Shimasaku et al. 2006) and $z\sim6.6$ (Taniguchi et al. 2005; Kashikawa et al. 2006). 

While the data accumulated on LAEs shows no evolution in the apparent Ly$\alpha$ luminosity function (LF) between $z=3.1$ - $5.7$ (Ouchi et al. 2008), the LF changes appreciably between $z=5.7$ and $6.6$ (Kashikawa et al. 2006) with $L_*$ at $z=6.6$ being about 50\% of the value at $z=5.7$. Surprisingly, however, the UV LF does not show any evolution between these same redshifts. Kashikawa et al. (2006), have proposed this evolution in the Ly$\alpha$ LF to be indicative of a sudden change in the ionization state of the Universe. However, the problem of why reionization would affect the high luminosity end of the LF rather than the the low, faint end, as expected, remains.

A range of theoretical models, both semi-analytic (Dijkstra et al. 2007a,b; Kobayashi et al. 2007,2009 ; Dayal, Ferrara \& Gallerani 2008 [DFG08]) and those involving cosmological simulations (McQuinn et al. 2007; Nagamine et al. 2008) have been built to explain the observations and use LAEs as probes of reionization. Using a search in a parameter space comprised by the SFR efficiency and Ly$\alpha$ transmission, Dijkstra et al. (2007b), showed that the evolution in the LF between $z=5.7$ and $6.6$ could be explained solely by an evolution of the underlying mass function. Kobayashi et al. (2007, 2009), used a semi-analytic galaxy formation model, including the effects of dust and outflows on the escape fraction of Ly$\alpha$ photons to explain the same. Nagamine et al. (2008), instead, have used the duty cycle scenario (the fraction of LAEs turned on at a  certain time) to explain the UV LF at $z=3$ and $6$. McQuinn et al. (2007), have shown that reionization increases the measured clustering of LAEs and hence, these objects are very useful to probe the epoch of reionization.

In DFG08 we showed that the Ly$\alpha$ and UV LFs can be explained by an evolution of the underlying dark matter halo mass function and dust attenuation that increases with decreasing redshift. We found that the ERM (Early Reionization Model), in which reionization ends at $z=7$, explains the data more consistently as compared to the LRM (Late Reionization Model, reionization ends at $z=6$). We thus found that reionization does not play any role in shaping the Ly$\alpha$ LF at $z\leq 6.6$. This is  consistent with the results obtained by McQuinn et al. (2007), who find that the Universe must be highly ionized at $z \sim 6.6$. In addition to the LFs, our model also reproduced the weighted skewness measurements (Kashikawa et al. 2006), the equivalent width (EW) at $z=4.5$ (Dawson et al. 2007) and predicted the SFR density of LAEs. However, both the average and distribution of EWs from our model at $z=5.7$ were much lower than those observed by Shimasaku et al. (2006).

In spite of these different approaches, there are still a number of missing ingredients, the most important of these being the calculation of the intrinsic Ly$\alpha$ luminosity, UV luminosity and SED (spectral energy distribution) as a function of the SFR, age, metallicity and IMF of the galaxy under consideration. Other important effects include the boost in the ionization rate imparted by galaxy clustering, its effects on the visibility of galaxies of different masses and the effects of inflows/outflows on the Ly$\alpha$ luminosity. Up to now, only a general value of this boost has been used (Dijkstra et al. 2007a); however, the effects of clustering on the visibility of LAEs as a function of galaxy properties have largely remained unexplored.


In this paper, we use state-of-the-art cosmological SPH simulations to fix the SFR, age, metallicity for each galaxy to obtain the intrinsic Ly$\alpha$ luminosity, UV luminosity and the SED. We again use the ERM (Gallerani et al. 2007), which accounts for all available data beyond LAEs, including Ly$\alpha$/Ly$\beta$ Gunn-Peterson opacity, electron scattering optical depths, Lyman limit systems, cosmic SFR histories and the number density of high redshift sources. Using the above ingredients, we obtain the Ly$\alpha$ and UV LFs and we are able to quantify the importance/effect of clustering on Ly$\alpha$ luminosity transmission and its contribution to shaping the Ly$\alpha$ LF. By doing so, we gain insight on the nature of LAEs and put constraints on their elusive physical properties.


\section{Simulations}
\label{simulations}
The simulation analyzed in this paper has been carried out using the
TreePM-SPH code {\small {GADGET-2}} (Springel 2005), with the
implementation of chemodynamics as described by Tornatore et
al. (2007). It is part of a larger set of cosmological runs, which are
presented and discussed in detail elsewhere (Tornatore et al. 2009, in
preparation). The adopted cosmological model corresponds to the
$\Lambda$CDM Universe with $\Omega_{\rm m }=0.26,\
\Omega_{\Lambda}=0.74,\ \Omega_{\rm b}=0.0413$, $n_s=0.95$, $H_0 = 73$
km s$^{-1}$ Mpc$^{-1}$ and $\sigma_8=0.8$, thus consistent with the
5-year analysis of the WMAP data (Komatsu et al. 2009). The
periodic simulation box has a comoving size of $75 h^{-1} {\rm Mpc}$
and contains $512^3$ Dark Matter particles and initially, an equal number
of gas particles.  As such, the masses of the DM and gas particles are
$m_{\rm DM}\simeq 1.7\times 10^8\,h^{-1}{\rm M}_\odot$ and $m_{\rm
  gas}\simeq 4.1\times 10^7\,h^{-1}{\rm M}_\odot$, respectively. The
Plummer--equivalent softening length for the gravitational force is
set to $\epsilon_{\rm Pl}=2.5\, h^{-1}$kpc, kept fixed in physical
units from $z=2$ to $z=0$, while being $\epsilon_{\rm Pl}=7.5\,
h^{-1}$kpc in comoving units at higher redshift. The value of the
softening parameter of the SPH kernel for the computation of
hydrodynamic forces is allowed to drop at most to half of the
the gravitational softening.

The run assumes a metallicity-dependent radiative cooling (Sutherland
\& Dopita 1993) and a uniform redshift-dependent Ultra Violet
Background (UVB) produced by quasars and galaxies as given by Haardt
\& Madau (1996). The code also includes an effective model to describe
star formation from a multi-phase interstellar medium (ISM) and a
prescription for galactic winds triggered by supernova (SN) explosions
(see Springel \& Hernquist 2003 for a detailed description). Galactic
winds are assumed to have a fixed velocity of 500 km s$^{-1}$, with a
mass upload rate equal to twice the local star formation rate.
The code includes the description of chemical enrichment given in Tornatore
et al. (2007). Metals are produced by SNII, SNIa and intermediate and
low-mass stars in the asymptotic giant branch (AGB).  We assume SNII
arise from stars having mass above $8\,{\rm M_\odot}$. As for SNIa, we
assume their progenitors to be binary systems, whose total mass lies
in the range (3--16)$\,{\rm M_\odot}$. The relative number of stars of
different mass is computed for this simulation by assuming the
Salpeter (1955) IMF between 1 and 100 ${\rm M_\odot}$.  
Metals and energy are released by stars of different masses by properly
accounting for mass--dependent lifetimes. In this work we assume the
lifetime function proposed by Padovani \& Matteucci (1993).  We adopt
the metallicity--dependent stellar yields from Woosley \& Weaver (1995) and
the yields for AGB and SNIa from van den Hoek \& Groenewegen (1997).

As for the identification of galaxies, they are recognized as
gravitationally bound groups of star particles.  For each analyzed
snapshot we first run a standard friends-of-friends (FOF) algorithm
with a linking length of 0.2 in units of the mean particle
separation. Each FOF group is then decomposed into a set of disjoint
substructures, which are identified by the SUBFIND algorithm (Springel
et al. 2001) as locally overdense regions in the density field of the
background main halo. After performing a gravitational unbinding
procedure, only sub-halos with at least 20 bound particles are
considered to be genuine structures (see Saro et al. 2006, for
further details). For each ``bona-fide'' galaxy, we compute the
mass-weighted age\footnote{This method tends to slightly bias the age towards larger values. On the other hand
numerical resolution limits the ability to resolve the smallest halos harboring the oldest stars.}, 
the total halo/stellar/gas mass, the SFR, the mass
weighted gas/stellar metallicity, the mass-weighted gas temperature
and the half mass radius of the dark matter halo.

We compute the Ly$\alpha$ emission and spectral properties for all the
structures identified as galaxies in the simulation boxes at the
redshifts of interest ($z\sim 5.7, 6.6, 7.6$). Obviously, not all
these galaxies will be necessarily classified as LAEs.

\section{Physics of Lyman Alpha emission}
\label{phy_of_lya}
We summarize here the main features of our model, the details of which can be found in DFG08. Star formation in galaxies gives rise to continuum and \HI ionizing photons, of which the the latter ionize the Inter Stellar Medium (ISM). Due to the high density of the ISM, recombinations take place on a short timescale and this gives rise to a Ly$\alpha$ emission line.

The intrinsic UV continuum and Ly$\alpha$ luminosity depend on the galaxy properties including the IMF, SFR ($\dot M_*$), stellar metallicity ($Z$) and age ($t_*$); all these quantities are taken from the simulation outputs, as discussed above. {\rm STARBURST99}, (Leitherer et al. 1999), a population synthesis code, is used to obtain the intrinsic values of the continuum luminosity as well as the rate of \HI ionizing photons produced using the Salpeter IMF, appropriate SFR, age and metallicity for each simulated galaxy. From our model, for a galaxy with $t_*=200 \, {\rm Myr}$, $Z = 0.2 Z_\odot$, $ \dot M_* = 1 {\rm M_\odot} \, {\rm yr^{-1}}$, the rate of production of \HI ionizing photons, $Q = 10^{53.47} {\rm s^{-1}}$ and the corresponding intrinsic Ly$\alpha$ luminosity is $L_\alpha^{int}=3.25 \times 10^{42} {\rm erg \, s^{-1}}$. This is very consistent with the value of $3.3 \times 10^{42} {\rm erg \, s^{-1}}$ shown in Tab. 4  of Schaerer (2003) for similar values of the age, metallicity and IMF. For the same galaxy, the intrinsic continuum luminosity is $L_c^{int} = 3.5 \times 10^{40} {\rm erg \, s^{-1}\AA^{-1}}$, which yields an EW of about $93$ \AA.


The Ly$\alpha$ line profile is affected by the rotation velocity of the galaxy ($v_c$), unlike the continuum.  For quiescent star formation, for realistic halo and disc properties, the galaxy rotation velocity can have values between 1-2 times the halo rotation velocity (Mo, Mao \& White 1998; Cole et al. 2000). We use a value of $1.5$ in this work. 
We calculate the velocity of the halo, $v_h$, assuming that the collapsed region has an overdensity of roughly 200 times the mean cosmic density contained in a radius $r_{200}$. Then, $v_h$, the velocity at $r_{200}$ is expressed as 
\begin{equation} 
v_h^2(z) ={\frac{G M_h}{r_{200}}} = G M_h \left[\frac{100 \Omega_m(z) H(z)^2}{G M_h}\right]^{1/3},  
\label{rot vel} 
\end{equation} 
where $M_h$ is the dark matter halo mass and $\Omega_m(z)$, $H(z)$ are the matter density and Hubble parameters, respectively, at the redshift of the emitter. 

Two processes determine the Ly$\alpha$ luminosity which emerges from the galaxy. First, only a fraction of the \HI ionizing photons ionize the ISM, contributing to the Ly$\alpha$ luminosity while the rest ($f_{esc}$) escape the galaxy and ionize the Inter Galactic Medium (IGM) surrounding it. Second, only a fraction, $f_\alpha$, of the Ly$\alpha$ photons produced escape the galaxy, unabsorbed by dust in the ISM.

The Doppler broadened Ly$\alpha$ luminosity profile that emerges from the galaxy is expressed as
\begin{equation}
L_\alpha^{em} (\nu) = \frac{2}{3} Q h \nu_\alpha (1-f_{esc}) f_\alpha \frac{1}{\sqrt{\pi} \Delta \nu_d} \exp^{-(\nu-\nu_\alpha)^2 / \Delta \nu_d^2} ,
\end{equation}
where $Q$ is the \HI ionizing photon production rate for the intrinsic galaxy properties, $h$ is Planck's constant, $\nu_\alpha$ is the frequency of Ly$\alpha$ photons, $\Delta \nu_d = [v_c/c] \nu_\alpha$, $c$ is the speed of light and the factor two-thirds arises assuming Case B recombination (Osterbrock 1989). 

The continuum band (1250-1550 \AA, centered at 1375 \AA) is chosen such that it is unaffected by attenuation due to \HI and it is only attenuated by dust in the ISM. The fraction, $f_c$ ($f_\alpha$) of continuum (Ly$\alpha$) photons which escape the galaxy depends on the total amount of dust in the ISM as well as its distribution (homogeneous or clumped).  The continuum luminosity emerging from the galaxy is expressed as $L_c^{em} = f_c L_c^{int}.$

Ly$\alpha$ photons suffer further attenuation as they travel through the IGM due to their large \HI optical depth; even small amounts of neutral hydrogen in the IGM can attenuate the Ly$\alpha$ luminosity by large amounts. To calculate this attenuation, we use the prescription detailed in Sec 2.4 of DFG08. We use the value of the UVB photoionization rate ($\Gamma_B$) from Gallerani et al. (2007) for the Early Reionization Model (ERM), wherein reionization ends at $z\sim7$, to calculate the mean neutral hydrogen fraction in the IGM\footnote{This is slightly inconsistent with the UVB used in the simulations. However, this produces negligible differences in the properties of the LAEs.}, $\chi_{HI} = n_{HI}/n_H$. In principle, our simulation could be used to account for inhomogeneities in the IGM along different lines of sight. However, this work is much beyond the scope of this paper and so, we assume a homogeneous and isotropic IGM density field. According to these calculations, at $z\sim(5.7,6.6,7.6)$, $\chi_{HI} = (6 \times 10^{-5}, 2.3 \times 10^{-4}, 0.16)$. For the given value of $\chi_{HI}$, we calculate the volume ($V_I$) of the so-called Str\"omgren sphere built by the \HI ionizing 
photons that escape the galaxy without ionizing the ISM and instead ionize the IGM (Shapiro \& Giroux, 1987; Madau, Haardt \& Rees, 1999) as:
\begin{equation}
\label{vstrom}
\frac{dV_I}{dt} - 3H(z) V_I = \frac{Q f_{esc}}{\chi_{HI} n_H} - \frac{V_I}{t_{rec}} ,
\end{equation}
where $n_H$ is the hydrogen number density at the redshift considered and $t_{rec}$ is the volume averaged recombination time in the IGM. 

Estimating $f_{esc}$ proves to be somewhat of a challenge. While empirical estimates from normal local and high redshift galaxies have yielded modest values of a few percent, theoretical studies have been largely inconclusive. Recently, using an N-body$+$ hydrodynamic simulation, Gnedin et al. (2008), have found an average value, $f_{esc} \sim 0.02$, for halo masses between $10^{10-12} {\rm M_\odot}$ over redshifts between 3 and 5. We use this value as a reasonable best estimate of $f_{esc}$ for all the galaxies in the simulation boxes at $z\sim5.7,6.6,7.6$ to calculate the size of their Str\"omgren Sphere. We could also calculate $f_{esc}$ by coupling our simulation with one that deals with radiative transfer. However, such a detailed analysis is deferred to future works.
 
Inside the Str\"omgren sphere, the ionization rate has two contributions : (a) a constant value, $\Gamma_B$, determined by the UVB photoionization rate  and (b) a radius dependent value, $\Gamma_L$, determined by the luminosity that emerges out from the galaxy. The total photoionization rate ($\Gamma_{BL}$) at any distance $r$ from the galaxy can be expressed as
\begin{eqnarray*}
\Gamma_{BL} (r) & = & \Gamma_B + \Gamma_L (r), \\
\Gamma_{BL} (r)& =  &\Gamma_B + \int _ {\nu_L} ^\infty \frac{L_\nu^{em}}{4 \pi r^2} \frac{\sigma_L}{h \nu} \bigg(\frac{\nu_L}{\nu}\bigg)^3 d\nu,
\end{eqnarray*}
where $L_\nu^{em}= L_\nu^{int} f_{esc} [\rm erg \, s^{-1} \, Hz^{-1}]$, is the specific ionizing luminosity emerging from the emitter, $\nu_L$ is the frequency corresponding to the Lyman limit wavelength (912 \AA) and $\sigma_L$ is the hydrogen photoionization cross-section. Since the IGM is in local photoionization equilibrium, we use ionization-recombination balance to compute the value of $\chi_{HI}$ at each point within the Str\"omgren Sphere. At the edge of this region, we force $\chi_{HI}$ to attain the constant global value, i.e., $\Gamma_{BL} \sim \Gamma_B$. 

If $z_{em}$ and $z_{obs}$ are the redshifts of the emitter and observer respectively, we calculate the total optical depth ($\tau_\alpha$) to the Ly$\alpha$ photons along the line of sight (LOS) as 
\begin{eqnarray}
\tau(\nu) &=& \int_{z_{em}}^{z_{obs}} \sigma\, n_{HI}(z)\, \frac{dl}{dz} dz ,\\
& = & \int_{z_{em}}^{z_{obs}} \sigma_0 \phi(\nu) n_{HI}(z) \frac{dl}{dz} dz ,
\end{eqnarray}
where $dl/dz = c/[H(z) (1+z)]$. Here, $\sigma$ is the total absorption cross-section and $\phi$ is the Voigt profile. We use $\sigma_0 = \pi e^2 f/(m_e c)$, where $e$, $m_e$ are the electron charge and mass respectively and $f$ is the oscillator strength (0.4162).

For regions of low \HI density, the natural line broadening is not very important and the Voigt profile can be approximated by the Gaussian core: 
\begin{equation} 
\phi(\nu) \equiv \phi_{gauss} = \frac{1}{\sqrt{\pi} (b/\lambda_\alpha)} e^{-({\frac{\nu_i-\nu_\alpha}{\nu_\alpha} \frac{c}{b}})^2},
\label{gauss} 
\end{equation} 
In eq. \ref{gauss}, $\nu_i$ is used since a photon of initial frequency $\nu$ has a frequency $\nu_i = \nu [(1+z_i)/(1+z_{em})]$ at a redshift $z_i$ along the LOS. The Doppler width is expressed as $b/\lambda_\alpha$, where $b=\sqrt {2kT/m_H}$ is the Doppler width parameter, $m_H$ is the hydrogen mass, $k$ is the Boltzmann constant and $T=10^4 K$ is the IGM temperature (Santos 2004; Schaye et al. 2000; Bolton \& Haehnelt 2007).
In more dense regions the Lorentzian damping wing of the Voigt profile becomes important.
According to Peebles (1993), this can be approximated as  
\begin{equation} 
\phi_{lorentz}(\nu_i) = \frac{\Lambda (\nu_i /\nu_\alpha)^4}{4\pi^2(\nu_i-\nu_\alpha)^2 + \frac{\Lambda^2}{4} (\nu_i/\nu_\alpha)^6} 
\end{equation} 
where $\Lambda= 6.25\times 10^8$~s$^{-1}$ is the decay constant for the Ly$\alpha$ resonance. 
Although computationally more expensive than the above approximations, using the Voigt profile to compute the absorption cross-section gives precise results, and therefore we have implemented it in our code to obtain all the results presented below.

The observed Ly$\alpha$ luminosity ($L_\alpha$) can be expressed as $L_\alpha = L_\alpha^{em} T_\alpha$ where a fraction $T_\alpha = e^{-\tau_\alpha}$ of the Ly$\alpha$ luminosity emerging from the galaxy is transmitted through the IGM.
Since the continuum is unaffected by transmission through the IGM, the observed continuum luminosity is calculated to be $L_c = L_c^{em}.$

\section{Effects of Clustering}
\label{clus}
One of the most important drawbacks of modelling LAEs semi-analytically is that only an average value of the UVB photoionization rate, $\Gamma_B$, is usually assumed, making any inference on the additional contribution, $\Gamma_G$, due to galaxy clustering not possible. As McQuinn et al. (2007), have shown, the observed clustering of LAEs is increased by transmission through a patchy IGM as expected during reionization. In addition to being an important tool to probe reionization, clustering will also leave imprints on the Ly$\alpha$ LF.

\begin{figure} 
  \center{\includegraphics[scale=0.5]{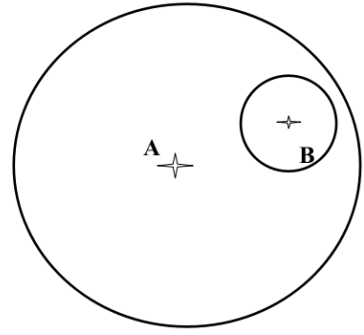}} 
  \caption{A schematic diagram to elucidate the effects of clustering on galaxy visibility (see text). The circles indicate the boundaries of individual \HII regions.}
\label{sch} 
\end{figure}  

Thanks to our cosmological SPH simulations, we are able to include the effect of clustering on the Ly$\alpha$ LF and transmission. Clustering of galaxies implies that the mean separation could become smaller than the typical size of their ionized regions. Therefore, more than one galaxy affects the size of the \HII region and the neutral hydrogen profile within it, thereby increasing the visibility of individual galaxies.
Consider for example the case of a luminous galaxy, A, and a fainter one, B (see Fig. 1). Each one of them will carve an \HII region in the IGM, whose size depends on their luminosity, as shown in the Figure. Depending on the IGM hydrogen neutral fraction, none, one or both galaxies can be seen individually. This is because, to a first approximation, the spatial scale imposed by the Gunn-Peterson damping wing on the size of the \HII region corresponds to a redshift separation of $\Delta z \approx 0.01$, i.e. about 200 kpc (physical) at $z=10$ (Miralda-Escud\'e 1998). However, when the mean separation between galaxies is smaller than the size of the smallest \HII region (i.e. galaxies are considerably clustered) then it can occur that both galaxies can be observed as LAEs due to the increased \HII region size and decreased \HI fraction within it boosting the transmissivity. The importance of such clustering effect on LAE visibility is stronger in the initial reionization phases, when the gas is almost neutral, implying that only clustered galaxies would be visible under those circumstances.

To quantify this effect from simulations, we use a post-processing technique in which we start by calculating the size of the ionized region around each galaxy as explained in eq. (\ref{vstrom}). Any two galaxies that are separated by a distance smaller than either of their Str\"omgren radii are then treated as a local enhancement of the photoionization rate for the other, i.e., for galaxy A, we add an extra contribution, $\Gamma_G$, evaluated as the local value of the photoionization rate of galaxy B at the position of galaxy A. The same procedure is followed when considering the transmissivity of galaxy B. More generally, the total ionization rate, $\Gamma_T$, seen at the position of galaxy $j$, whose separation from $N$ other galaxies is smaller than either of the Str\"omgren radii can be expressed as
\begin{equation}
\Gamma_{T,j} (r) = \Gamma_{BL}(r) + \sum_{i=1, i \neq j} ^ N \int_{\nu_L}^\infty \frac{L_{\nu,i}^{em}}{4 \pi r_{ij}^2} \frac{\sigma_L}{h \nu} \bigg(\frac{\nu_L}{\nu}\bigg)^3  d\nu, 
\end{equation}
where the second term on the right hand side represents $\Gamma_G$. Further, $L_{\nu,i}^{em}$ is the ionizing luminosity emerging from the $i^{th}$ galaxy and $r_{ij}$ is the radial distance between galaxies $i$ and $j$.
As a caveat, we point out that we are calculating clustering effects assuming that the sizes of the \HII regions correspond to sources embedded in the IGM at the mean
ionization fraction given by the ERM. However, prior to complete overlap, the ionization field is very patchy and galaxies are more likely to be immersed in either an almost neutral or a highly ionized region. Because the average $\chi_{HI}=0.16$, it means that $1-\chi_{HI}=84$\% of the volume is substantially ionized at $z=7.6$. Hence, in general, our method provides a lower limit to the number of detectable LAEs. Several authors (Zahn et al. 2007; Mesinger \& Furlanetto 2007; Geil \& Wyithe 2008) have presented schemes that avoid detailed radiative transfer (RT) calculations and still provide ionization schemes in good agreement with simulations. A precise calculation of clustering effects can also be done by properly following radiative transfer in detail. We defer these inclusions to our model to further work.

To summarize, the boost in the ionization background imparted by clustering is important for all galaxies when the IGM is close to neutral. The importance of this effect decreases as reionization proceeds, however, it decreases faster for more luminous galaxies.

\section{Results}
\label{results}

We are now ready to compare model results with observations. In particular, we compare the calculated UV and Ly$\alpha$ LFs to the data obtained by Shimasaku et al. (2006) ($z\sim5.7$) and Kashikawa et al. (2006) ($z\sim6.6$). We also predict the LFs for $z\sim7.6$ although data are not yet available at this redshift. In addition, we also explore the effects of clustering on Ly$\alpha$ transmission and present synthetic spectral energy distribution (SEDs) to match with Lai et al. (2007) observations at $z\sim5.7$. Finally, we will discuss the physical properties such as age, metallicity, halo/stellar mass and SFR for all objects identified as LAEs from this work.

\subsection{Ly$\alpha$ LF and the effects of clustering}
\label{lya}

\begin{figure} 
  \center{\includegraphics[scale=0.5]{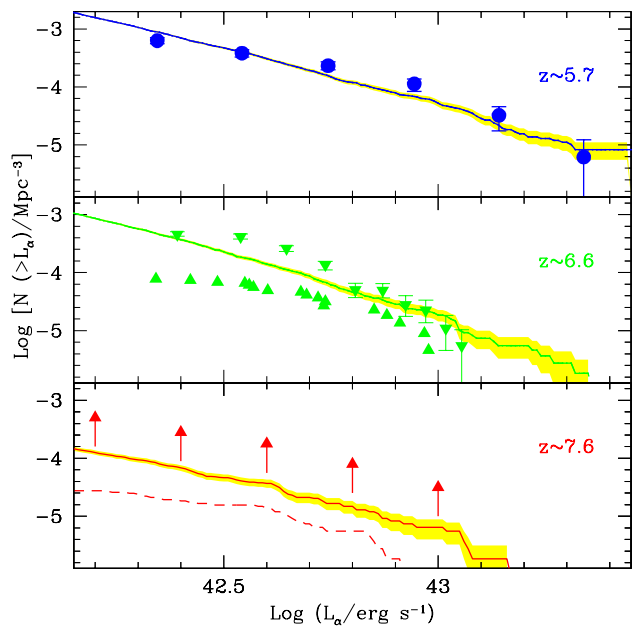}} 
\caption{Cumulative Ly$\alpha$ LF for the ERM. The panels are for $z\sim 5.7$, $6.6$ and $7.6$ from top to bottom. Points represent the data at two different redshifts:  $z\sim 5.7$ (Shimasaku et al. 2006) (circles), $z \sim 6.6$ (Kashikawa et al. 2006) with downward (upward) triangles showing the upper (lower) limits. Solid (dashed) lines in each panel refer to model predictions at $z\sim 5.7,6.6,7.6$ for the parameter values in Tab. \ref{table1} including (excluding) clustering effects. Shaded regions in all panels show poissonian errors. Though curves with/without clustering effects are indistinguishable at $z=5.6, 6.6$; the difference is appreciable at $z=7.6$. Because radiative transfer is not accounted for in the simulations, at $z=7.6$ our results represent lower limits to the observed objects.}
\label{lya3} 
\end{figure}

As mentioned before, we use the procedure explained in Sec. \ref{phy_of_lya} to calculate the observed Ly$\alpha$ luminosity for each of the galaxies in the simulation boxes. Galaxies with (a) observed Ly$\alpha$ luminosity in the currently observable range, $L_\alpha \geq 10^{42.2} {\rm erg \, s^{-1}}$ and (b) value of the observed EW, $EW>20$ \AA\, are then identified as LAEs, which are used to build the cumulative Ly$\alpha$ LF to compare to observations. The number of objects identified as LAEs from our simulations are (1696, 929, 136) at $z\sim (5.7, 6.6, 7.6)$; since the definition of LAE is an operational one based on the observed Ly$\alpha$ luminosity, the number depends on the adopted values of $f_\alpha$ and $T_\alpha$. The values given above are for the best fit parameters shown in Tab. 1 and Fig. \ref{lya3}. 

We find that, independent of clustering, to match the data at both $z\sim5.7$ and $6.6$, only a certain fraction ($f_\alpha = 0.3$) of the Ly$\alpha$ luminosity {\it must emerge out of the galaxy, unabsorbed by dust within the ISM}. As shown in the uppermost panel of Fig. \ref{lya3}, within error bars, the theoretical LF nicely matches the data at $z\sim 5.7$. 

\begin{table*} 
\begin{center} 
\caption {Best fit parameter values to match the Ly$\alpha$ and UV LFs in the ERM, including clustering effects. For each redshift (col. 1), we report the UVB photoionization rate (col. 2), the fraction of neutral hydrogen corresponding to the UVB photoionization rate (col. 3), the average value of the boost parameter (col. 4), the average transmission of the Ly$\alpha$ luminosity (col. 5), the value of the the escape fraction of Ly$\alpha$ photons (col. 6), the escape fraction of UV continuum photons (col. 7) and the color excess calculated using the supernova dust extinction curve (col. 8).}
\begin{tabular}{|c|c|c|c|c|c|c|c}
\hline 
$z$& $\Gamma_B$ & $\chi_{HI}$ &  $\langle B \rangle$ & $\langle T_\alpha \rangle$ &$f_\alpha$&$f_c$&$E(B-V)$\\  
$$& $(10^{-12} {\rm s^{-1}})$ & $$ &  $$ & $$ &$$ &$$ &$$\\  
\hline 
$5.7$& $0.47$ &$6 \times 10^{-5}$ & $1.04$ & $0.49$& $0.3$ & $0.22$ & $0.15$\\
$6.6$& $0.19$ & $2.3 \times 10^{-4}$ & $1.05$ & $0.49$ &$0.3$ & $0.37$ & $0.1$\\
$7.6$& $0.28\times 10^{-3}$ &$0.16$ & $58.5$ & $0.42$ & $0.3$ & $0.37$ & $0.1$\\
  \hline
\label{table1} 
\end{tabular} 
\end{center}
\end{table*}

With the same value of $f_\alpha=0.3$ at $z \sim 6.6$, the theoretical LF lies close to the upper bound of the data as shown in the central panel. Since the dust attenuation of Ly$\alpha$ photons does not evolve between $z \sim 6.6$ and $5.7$, it is reasonable to assume that the same value also holds at $z \sim 7.6$ (the cosmic time between $z \sim 7.6$ and $6.6$ is only about $0.1$ Gyr, which might be too small for any significant dust evolution) and then use it to predict the LF (bottom panel of Fig. \ref{lya3}). Fig. \ref{lya3} provides interesting information on the effects of clustering as well: from the overlapping of the LFs including/excluding clustering at $z \sim 5.7$ and $6.6$, we conclude that these are negligible at these epochs. Clustering of sources, however, plays a key role at $z \sim 7.6$. If clustering effects are neglected, very few objects (about 30) would be luminous enough to be detectable in current Ly$\alpha$ surveys; instead, the luminosity boost due to clustered sources leads to about 136 objects  to become visible in our simulation volume. We reiterate that, because we do not follow the radiative transfer of ionizing radiation through the IGM, the results presented in Fig. 2 (and in Fig. 4) at $z=7.6$ must be seen as lower limits to the actual number density of LAEs.    

We pause to discuss an issue concerning the LF at $z\sim 6.6$. The model predictions for the faint end of the LF lie between the upper and lower limits set by the observational data. Although this could be regarded as a success of the model, it is safe to discuss if physics not included in our model or biases in the data might spoil this agreement. While the lower bound of the data is made up of spectroscopically confirmed LAEs, the upper bound is a photometric sample composed of all galaxies identified as LAE candidates at this redshift. Spectroscopic analysis of LAE candidates on the upper bound of the LF could then possibly rule out a number of them as being low-$z$ interlopers and contaminants, thus bringing the upper limit in agreement with our curve. This would require that up to 60\% of the candidates might not be confirmed. Another possibility is that most of the faint objects are not individual galaxies but unresolved groups, a point made by Mori \& Umemura (2006), who have shown that a number of small galaxies undergoing mergers can be identified as a single LAE at high redshifts. At $z \sim 6.6$, the resolution of the data is about $1''$, which corresponds to a physical separation of $5.4$ kpc. Although in our simulations, we do not find any objects that are separated by such small distances, this might also be due to insufficient resolution on scales of about 5 kpc. In this work, we discard this possibility, since, although a small number of such pairs might be found with higher resolution, they would probably not be enough to boost up the low luminosity end of the LF substantially. 

\begin{figure} 
  \center{\includegraphics[scale=0.5]{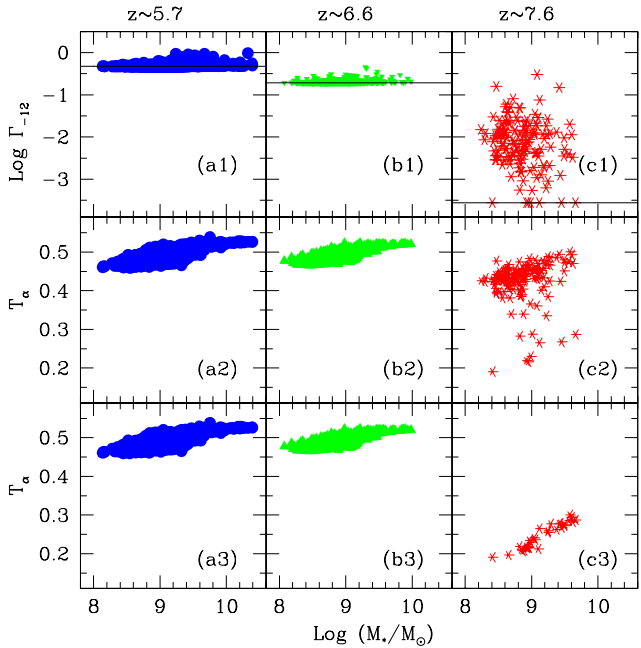}} 
  \caption{{\it Upper row}: Photoionization rates as a function of LAE stellar mass, $M_*$. The horizontal line corresponds to the contribution of the UVB, $\Gamma_B$; symbols in each panel denote the value of $\Gamma_B+\Gamma_G$. {\it Middle}: Ly$\alpha$ transmissivity, $T_\alpha$ including clustering effects. {\it Lower}: Ly$\alpha$ transmissivity without clustering effects. Columns refer to $z\sim 5.7,6.6,7.6$ as indicated.}
\label{clusfig} 
\end{figure}

We now come back to clustering. The reason why its effects become important at $z\sim 7.6$ is easily explained using Fig. \ref{clusfig}. First, from the uppermost three panels (a1, b1, c1), it is seen that the UVB photoionization rate ($\Gamma_B$) is very low at $z \sim 7.6$ ($2.8\times 10^{-16}$ s$^{-1}$). Between $z \sim 7.6$ and $6.6$, it increases rapidly by about 3 orders of magnitude such that at $z \sim 6.6$, $\Gamma_B = 1.9 \times 10^{-13}$ s$^{-1}$. Afterwards, the ionization rate increases only by a factor of about $2.5$ between $z \sim 6.6$ and $5.7$. Second, at $z \sim 5.7, 6.6$ the value of the boost parameter, $\langle B \rangle =  1 + \langle \Gamma_G \rangle / \Gamma_B$,\footnote{$\langle \Gamma_G \rangle = (1/N) \sum_{i=1}^N \Gamma_G(i)$, where $\Gamma_G(i)$ is the photoionization boost due to clustered LAEs seen by the $i^{th}$ LAE of the total $N$ LAEs at the redshift considered.} is less than a factor of 1.5, while at $z \sim 7.6$, $\langle B \rangle \approx 58$, i.e. the photoionization rate is strongly dominated by the local emission from the clustered LAEs. Since the IGM is already highly ionized at $z < 7$ in the ERM, the extra local contribution from clustered LAEs does not affect the Ly$\alpha$ transmission sensibly, as seen from the comparison of panels (a2,a3) and (b2,b3) of Fig. \ref{clusfig}. However, at $z \sim 7.6$, the effects of clustering on the transmissivity, $T_\alpha$, are dramatic. The photoionization rate boost due to clustered LAEs makes the IGM transparent enough that about 136 in our simulation volume become visible as compared to 30 that would be detected as LAEs in the absence of clustering effects. This is clear from the comparison of panels (c2) and (c3), from which we conclude that the transmissivity is increased up to values of $T_\alpha=0.2-0.5$ when clustering is included in the computation as compared to $T_\alpha=0.2-0.3$ when it is not.

A few additional points are worth mentioning here: (a) on a galaxy to galaxy basis, it is not necessarily the environments of the most massive galaxies that experience the largest enhancement of the photoionization rate due to clustering as seen from the upper three panels of Fig. \ref{clusfig}. This shows that the boost in photoionization rate seen by a small galaxy within the \HII region of a more massive one is greater than or comparable to the boost seen by a massive galaxy due to the contribution from a large number of small galaxies embedded in its \HII region. (b) If $\chi_{HI}\ll 1$, both including/excluding the contribution of $\Gamma_G$, $T_\alpha$ increases with $M_*$ and hence with star formation rate (panels a2, a3, b2, b3). This leads us to conclude that for a highly ionized IGM, the contribution of $\Gamma_L (\propto \dot M_*)$ always dominates over that of $\Gamma_G$. (c) If $\chi_{HI} \approx 1$, $T_\alpha$ still increases with the star formation rate both including/excluding $\Gamma_G$, although the scatter is much larger in the case including $\Gamma_G$. This is because excluding $\Gamma_G$, very few galaxies are luminous enough to transmit enough of the Ly$\alpha$ luminosity to be visible as LAEs (panel c3). However, including $\Gamma_G$ dramatically increases the $T_\alpha$ such that about 4 times as many galaxies become visible (panel c2). Hence, we conclude that in this case, $\Gamma_G$ contributes significantly in making the environment around LAEs more transparent to Ly$\alpha$ photons.


\subsection{UV LF}
\label{uvlf}

\begin{figure} 
  \center{\includegraphics[scale=0.5]{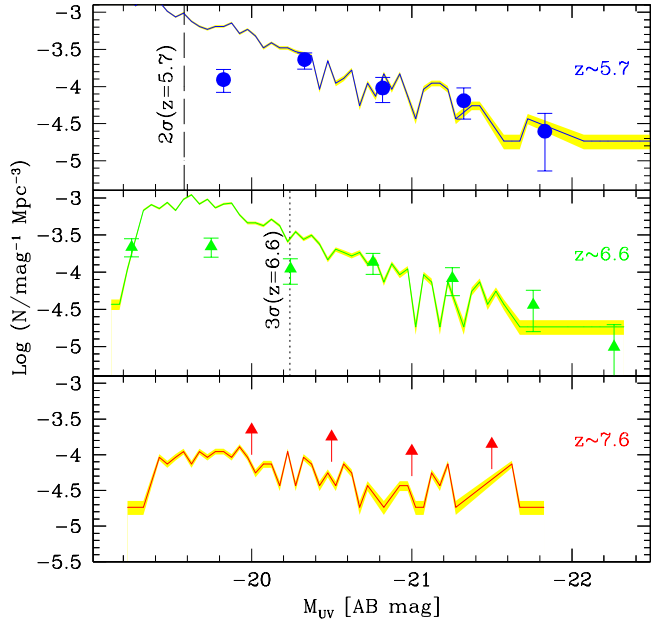}} 
  \caption{UV LAE LF for the ERM. Points represent the data at two different redshifts: $z \sim 5.7$ (Shimasaku et al. 2006) (circles), $z \sim 6.6$ (Kashikawa et al. 2006) (triangles). Lines refer to model predictions including clustering at the redshifts (from top to bottom): $z \sim 5.7,6.6,7.6$, for the parameter values in Tab. \ref{table1}. The vertical dashed (dotted) lines represent the observational 2$\sigma$ (3$\sigma$) limiting magnitude for $z = 5.7$ ($z =6.6$). The shaded region in all panels shows the poissonian errors. Because radiative transfer is not accounted for in the simulations, at $z=7.6$ our
results represent lower limits to the observed objects.} 
\label{uvlf3} 
\end{figure} 

For each galaxy identified as a LAE included in the Ly$\alpha$ LF, we calculate the total continuum luminosity in a band between 1250-1500 \AA, centered at 1375 \AA. We bin the number of galaxies on the basis of the continuum luminosity magnitude and divide by the volume of the simulation box ($75^3 h^{-3}$ comoving ${\rm Mpc^3}$) to obtain the UV LF for the LAEs identified in Sec. \ref{lya}. This is shown in Fig. \ref{uvlf3}. 

We find that only a certain fraction of the continuum photons, $f_c$, must escape the galaxy undamped by dust in the ISM to match to the observations; this fraction is found to decrease with decreasing redshift, going from  $0.37$ at $z \sim 6.6$ to $0.22$ at $z \sim 5.7$. This means that while 37\% of all continuum photons escape the galaxy at $z \sim 6.6$, only $22$\% escape at $z \sim 5.7$. We interpret this decreasing $f_c$ to be the result of an increase in the dust content of the galaxies. To make predictions at $z \sim 7.6$, we again use the same value of $f_c$ as at $z \sim 6.6$, making the assumption that the dust content of LAEs does not evolve between these two redshifts. The values of $f_c$ for all the three redshifts are shown in Tab. \ref{table1}. 

\begin{figure*} 
  \center{\includegraphics[scale=1.0]{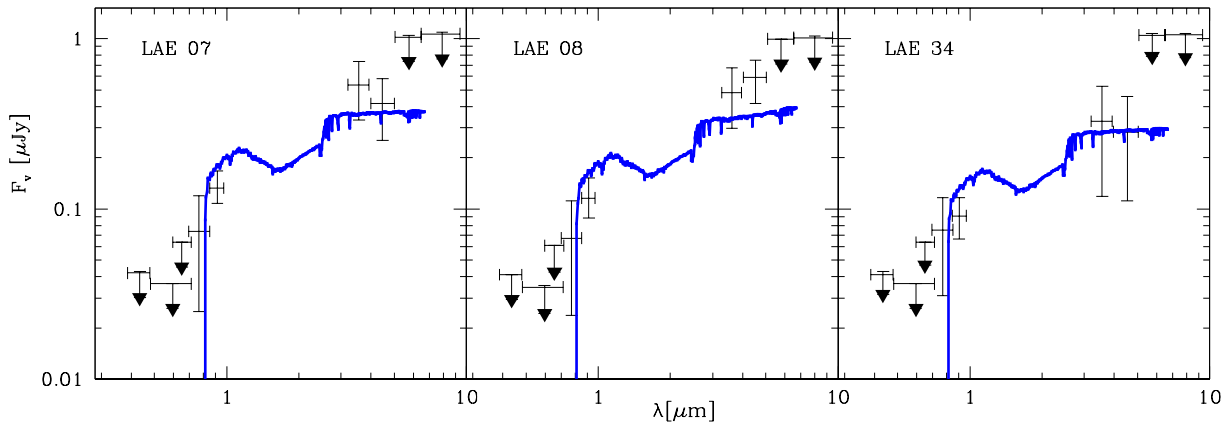}} 
  \caption{Comparison of theoretical SEDs (lines) with observations (points) for LAE \#07, \#08 and \#34,
(from left to right) from Lai et al. (2007). Points with downward pointing arrows represent the $3 \sigma$ upper limits of the data.}
\label{spec} 
\end{figure*}   

An interesting point here is that at $z \sim (5.7, 6.6)$, $f_\alpha/f_c \sim (1.4, 0.8)$, confirming a result obtained from our previous semi-analytical model, presented in DFG08. Using the Supernova extinction curve, $f_\alpha/f_c \sim 0.8$ for a homogeneous distribution of dust, Bianchi \& Schneider (2007). However, no single extinction curve (Galactic, Small Magellanic Cloud or Supernova) can give a value of $f_\alpha/f_c >1$. The relative damping at $z \sim 5.7$ can only be explained by an inhomogeneous two-phase ISM model in which clumped dust is embedded in a hot ISM, as proposed by Neufeld (1991). The data thus, seem to hint at the fact that the dust distribution in the ISM of LAEs becomes progressively inhomogeneous/clumped with decreasing redshift.

We calculate the color excess $E(B-V)$ ($ = A_v/R_v$) for each of the redshifts as a sanity check. Using the supernova dust extinction curve (Bianchi \& Schneider 2007), we calculate $A_v = A_\lambda (1375\, {\rm \AA}) /  5.38 $ and $R_v=2.06$. The color excess is then calculated as
\begin{equation}
E(B-V)=\frac{A_\lambda (1375\,{\rm \AA})}{5.38 R_v} = - \frac{2.5 \log_{10} f_c}{5.38 R_v}.  \\
\end{equation}
With the above formulation, we find that $E(B-V) \sim 0.15$ at $z \sim 5.7$, $E(B-V) \sim 0.1$ at $z \sim 6.6$. These finding are consistent with other independent data as well: in fact, Lai et al. (2007) have shown that fitting the SEDs of 3 LAEs observed by them at $z \sim 5.7$ requires $E(B-V) < 0.425\, (0.225)$ for young (old) galaxies.

\subsection{SEDs}
\label{sed}

Lai et al. (2007) have observed the spectra for three LAEs (\#07, \#08 and \#34) at $z \sim 5.7$ with observed Ly$\alpha$ luminosities, $L_\alpha = (4.9,  4.3, 3.6) \times 10^{42}\, {\rm erg\, s^{-1}}$ (K. Lai, private communication). From amongst the galaxies we identify as LAEs at $z\sim 5.7$, we select those three galaxies whose observed Ly$\alpha$ luminosities match most closely with the three values observed. The spectra of each of the selected galaxies (obtained using {\rm STARBURST99}, as explained in Sec. \ref{phy_of_lya}) is then attenuated using the SN dust extinction curve and  $E(B-V)=0.15$, to produce synthetic spectra which are shown in Fig. \ref{spec}. 

As shown by Lai et al. (2007), the spectra can be well fit by different kinds of stellar populations, varying the age, metallicity and color excess. However, we have no free parameters since the age, metallicity and SFR for each of the galaxies is obtained from the simulation outputs and the color excess value is obtained using the UV LF at $z \sim 5.7$ (Shimasaku et al. 2006) as mentioned before. As seen, a remarkable agreement is found between the synthetic and observed SEDs. Even though we are using a small and biased data set of LAEs selected based on the IRAC $3.6 \mu m$ and $4.5 \mu m$ detections, the agreement between the synthetic and observed SEDs provides a strong consistency test of our model.

We briefly mention the physical properties of the LAEs whose synthetic spectra are shown in Fig. \ref{spec}. The stellar ages  are 182--220 Myr for the three LAEs (see also Tab. \ref{tab_spec}). Hence, these LAEs are intermediate age objects, rather than being very old ($t_* \sim 700$ Myr) or very young ($t_* \sim 5$ Myr). Their stellar metallicities are about $0.2-0.3\, Z_\odot$, and the SFR are between 7-10 ${\rm M_\odot} {\rm yr^{-1}}$. Further, the SEDs for all the three LAEs are well reproduced by a single value of the color excess which shows that all these objects possibly contain similar amounts of dust in the ISM.

\begin{table} 
\begin{center} 
\caption {Physical properties of the LAEs from the simulation which match the observed Ly$\alpha$ luminosities most closely. For each LAE observed by Lai et al. (2007) (col. 1), we show the age (col. 2), stellar metallicity (col. 3), SFR (col. 4) and color excess (col. 5) for the corresponding LAE from our simulation.}
\begin{tabular}{|c|c|c|c|c} 
\hline 
$\# {\rm LAE} $ & $t_* $ & $Z $ & $\dot M_* $ & ${\rm E(B-V)}$  \\  
$ $ & ${\rm (Myr)}$ & $ (Z_\odot)$ & $ ({\rm M_\odot} {\rm yr^{-1}})$ & $$  \\  
\hline
$07$ & $191$ & $0.23$ & $9.7$ & $0.15$ \\
$08$ & $182$ & $0.32$ & $9.6$ & $0.15$ \\ 
$34$ & $220$ & $0.23$ & $7.3$ & $0.15$ \\  
 \hline
\label{tab_spec} 
\end{tabular} 
\end{center}
\end{table}

\subsection{The nature of LAEs}
\label{nature}

\begin{figure} 
\center{\includegraphics[scale=0.5]{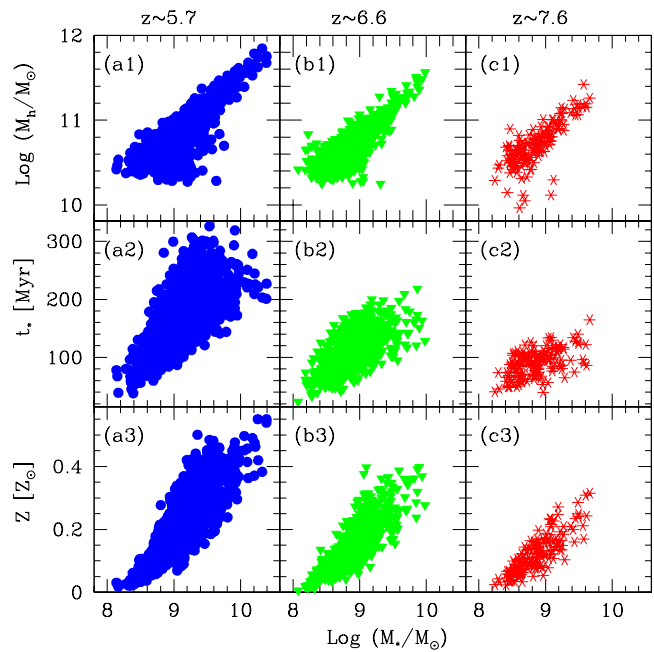}} 
\caption{Physical properties of LAEs at $z\sim 5.7$, $z \sim 6.6$ and $z \sim 7.6$ as referred to by columns. As a function of stellar 
mass, $M_*$, we show  (a) the halo mass, $M_h$, (a1-c1), (b)  mass-weighted stellar ages, $t_*$, (a2-c2), and (c) mass-weighted stellar 
metallicity, $Z$, (a3-c3).}
\label{phy1} 
\end{figure}   

Having selected the galaxies in the simulated volume that would be experimentally defined as LAEs, and having shown
that their luminosity function and SED are consistent with data, we can take a further step and quantify the physical
properties of these objects. This includes the relation between stellar and halo mass, age, metallicity, star formation rate and evolution. We will present these results up to $z \sim 7.6$; however it must be kept in mind that at that redshift, only 136 galaxies are bright enough to be observed using standard narrow-band techniques which makes the scatter much larger as compared to the lower redshifts.   

\begin{table*} 
\begin{center} 
\caption {For all the LAEs comprising the Ly$\alpha$ LF including clustering effects, at the redshifts shown (col. 1), we show the range of mass weighted ages (col. 2), the average stellar metallicity (col. 3), the average SFR (col. 4), the average intrinsic and observed EWs (col. 5, 6), the average value of the star formation indicator (col. 7) and the SFR density (col. 8) obtained from our simulation.}
\begin{tabular}{|c|c|c|c|c|c|c|c} 
\hline 
$z$ & $t_* $ & $\langle Z \rangle$ & $\langle \dot M_* \rangle $ & $\langle EW^{int}\rangle$ & $\langle EW \rangle $ & $\langle \cal I \rangle$ & $\dot\rho_* $  \\ 
$$ & $({\rm Myr})$ & $(Z_\odot)$ & $({\rm M_\odot} {\rm yr^{-1}})$ & $({\rm\AA})$ & $({\rm\AA})$ & $$ & $({\rm M_\odot} {\rm yr^{-1} Mpc^{-3}})$  \\ 
\hline
$5.7$ & $38-326$ & $0.22$ & $6.9$ & $94.3$ & $63.5$ & $0.72$ & $1.1 \times 10^{-2}$ \\
$6.6$ & $23-218$ & $0.15$ & $5.7$ & $104.1$ & $41.7$ & $0.72$ & $4.9 \times 10^{-3}$ \\ 
$7.6$ & $39-165$ & $0.12$ & $7.7$ & $108.9$ & $37.8$ & $0.78$ & $9.6 \times 10^{-4}$\\  
 \hline
\label{table2} 
\end{tabular} 
\end{center}
\end{table*} 

The first set of relations among stellar and halo mass, age, metallicity is presented in Fig. 6. LAEs are characterized by dark matter halo masses in the range $10^{10.2-11.8} M_\odot$ at $=5.7$, corresponding to $> 2\sigma$ fluctuations at all redshifts. This range becomes progressively narrower at earlier epochs because of two occurrences:
(i) the mass function of halos shifts to lower masses in hierarchical structure formation models; and (ii) smaller halos progressively become invisible at higher redshifts as their stellar mass and hence luminosity is too low to be detected (we recall that all the objects analyzed are part of the LFs shown in Figs. \ref{lya3} and \ref{uvlf3}). In spite of this fact, there is almost no evidence of evolution of the stellar to halo relation in the redshift range under examination, and we find $M_* \propto M_h^{1.56}$, with the best fit given by the following expression: 
\begin{equation}
\log_{10} (M_h) = (0.64-0.06\Delta z) \log_{10} (M_*)  + (5.0+0.5\Delta z), 
\end{equation}
where $\Delta z = (z-5.7)$. The above relation implies that the stellar mass per unit halo mass is increasing towards larger systems, which is just a restatement of the well-known fact that the star formation is less efficient in small galaxies due to the inhibiting effects of mechanical feedback. The typical stellar masses of LAEs are $< 10^{10.5}$ at all redshifts, i.e. they are smaller than the Milky Way.

\begin{figure} 
\center{\includegraphics[scale=0.5]{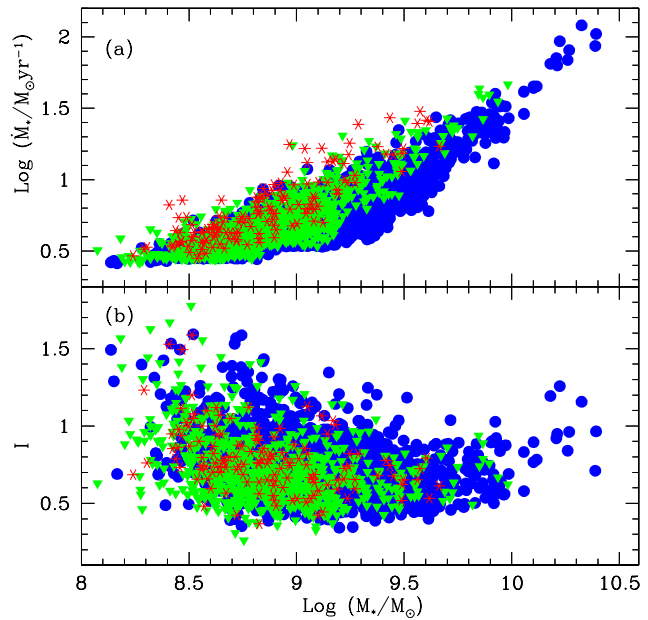}} 
\caption{(a) SFR, $\dot M_*$, and (b) SFR indicator, $\cal I$ = $\dot M_* t_*/M_*$, as a function of stellar mass ($M_*$) for LAEs at three redshifts: $z\sim 5.7$ (circles), $z\sim 6.6$ (triangles) and $z \sim 7.6$ (asterisks).}
\label{sfr} 
\end{figure}  

\begin{figure*} 
  \center{\includegraphics[scale=0.75]{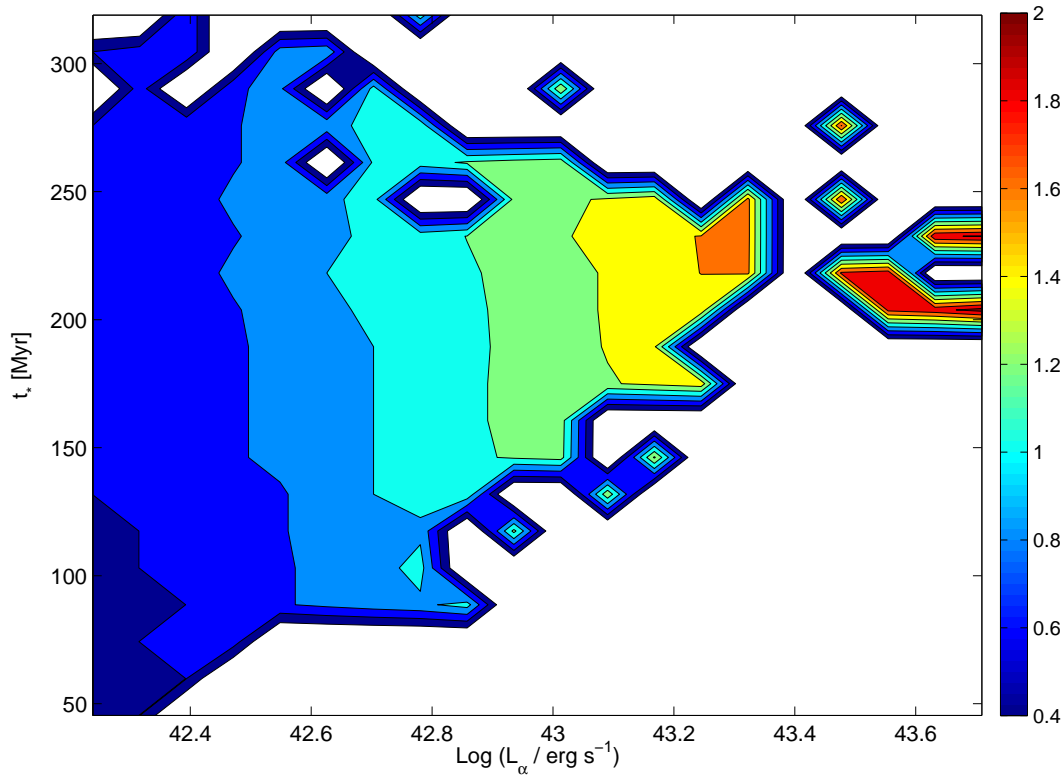}} 
  \caption{Contour plot showing relation between observed Ly$\alpha$ luminosity, age and SFR at $z\sim 5.7$. The SFR values, ${\rm Log} (\dot M_*/{\rm M_\odot} \, {\rm yr^{-1}})$, are color-coded by the bar on the right. }
\label{contour} 
\end{figure*}

Although a considerable spread is present, on average larger systems tend to be older than small ones. LAEs result from the merging of several sub-halo systems in which star formation was ignited long before. In Tab. \ref{table2}, we show that the range of ages reduces towards high redshifts from $38-326$ Myr at $z \sim 5.7$ to $39-165$ Myr at $z \sim 7.6$. These ages imply that the oldest stars in these systems formed already at $z=8.5$, thus during the reionization epoch. We reiterate that, because of finite numerical resolution, stars in halos below our resolution limit might have 
formed even before that epoch, so the previous value must be seen as a lower limit to the onset of star formation activity. From the previous discussion it is clear that LAEs are neither pristine nor very young objects forming their stars for the first time. This is confirmed by the range of metallicities found in these systems and shown in Fig. \ref{phy1} (see also Tab. \ref{table2}). Metallicities are in the range $0.02-0.55$ $Z_\odot$, with the mean over the sample decreasing with redshift; also, larger objects have higher metallicities as expected from their larger stellar masses. These results are broadly consistent with an enrichment predominantly caused by SNII, and therefore tightly following
the SFR of the galaxy; the increasing scatter seen at the lowest redshift hints at a larger contribution by SNIa. The mass-metallicity relation for LAEs is also an interesting outcome of our study. This can be conveniently expressed as 
\begin{equation}
Z / Z_\odot = (0.25 - 0.05\Delta z)\log_{10} (M_*)  - (2.0-0.3\Delta z),$$
\end{equation}
where $\Delta z = (z-5.7)$. Differently to the analogous relation observed at lower redshifts (Tremonti et al. 2004; Panter et al. 2008; Maiolino et al. 2008), we do not see the sign of a flattening of metallicity towards larger masses. As the flattening is usually interpreted as a result of a transition from a galactic wind-regulated metal
budget to a close-box evolution in which all metals are retained, we conclude that in our LAEs, winds play an important role because of their relatively low masses, a point already noticed above. 

We now turn to the analysis of the star formation properties of LAEs (Fig. \ref{sfr}). Star formation rates are in 
the range from 2.5-120 ${\rm M_\odot}\,{\rm yr^{-1}}$, i.e. a sustained but not exceptionally large star formation activity, with large objects at later times being the most prominent star factories. The relation appears to flatten below $M_* \leq 10^{9.3} {\rm M_\odot}$, as is clearly seen. Stated differently, low mass LAEs have star formation rates confined in the narrow range 2.5-10 ${\rm M_\odot}\,{\rm yr^{-1}}$, whereas only (relatively few) larger objects undergo intense star formation events, $\dot M_* > 50 \,{\rm M_\odot} \,{\rm yr^{-1}}$. This might be the result of feedback regulation, which prevents small objects from burning most of their gas fuel at high rates. To investigate this aspect more, we have studied the behavior of a star formation indicator, ${\cal I}=\dot M_* t_*/M_*$. Physically, this is the ratio between the stellar mass produced if a LAE had always formed stars at the rate deduced at the given redshift for a time equal to the mean age of its stars, and the actual total stellar mass of the system. Hence if ${\cal I} > 1$ (${\cal I} < 1$), the star formation rate was lower (higher) in the past. The large majority of LAEs show values of ${\cal I} < 1$, indicating that the star formation rate averaged over the entire history must be larger than the final value. This could be due to huge bursts of star formation, catalyzed by mergers. Some systems (at the low and high ends of the stellar mass distribution at $z \sim 5.7$) have ${\cal I} > 1$, i.e,  they have quietly built up their stellar population at an increasing rate as they grow in mass by subsequent mergings.  
We provide a handy fit for the SFR in terms of the stellar mass:
\begin{eqnarray*}
\log_{10} (\dot M_*) & = & (0.4-0.04\Delta z) [\log_{10} (M_*)]^2 -  \\
& & (6.9-0.8\Delta z ) \log_{10} (M_*) + (30-5\Delta z),
\end{eqnarray*}
where $\Delta z = (z-5.7)$. 

Finally, we discuss the dependence of the observed Ly$\alpha$ luminosity on age and SFR (Fig. \ref{contour}). 
The SFR for all galaxies with a given $L_\alpha$ are very similar, even though the ages 
vary between $40-340$ Myr.  As explained before, this is because according to the ERM, at $z \sim 5.7$, 
the Universe is already so ionized ($\chi_{HI} \sim 10^{-5}$) that even the smallest emitters 
are able to build large enough Str\"omgren spheres on short timescales. Further, the 
number of galaxies with small SFR ($\dot M_* \leq 20 \,{\rm M_\odot} {\rm yr^{-1}}$) is quite high, 
after which this number decreases rapidly (see also Fig. \ref{sfr}), with very few galaxies in the 
highest luminosity bins. Thus, the faint end of the LF samples a mixture 
of young and old objects, while LAEs in the bright end are predominantly massive, 
intermediate (200-250 Myr) age systems.

\section{Discussion}
\label{discussion}

\begin{figure} 
  \center{\includegraphics[scale=0.5]{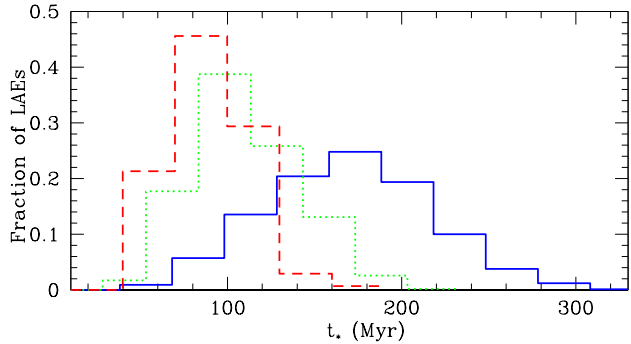}} 
  \caption{Normalized distribution of number of LAEs as a function of the mass weighted stellar age ($t_*$). The lines are for z: 5.7 (solid), 6.6 (dotted) and 7.6 (dashed).  }
\label{pdf} 
\end{figure}

The ages of LAE stellar populations are currently hotly debated, hence we need to discuss
our results in the framework of the various arguments given in the literature. For example,
Finkelstein et al. (2009) find a bimodality in their sample of 14 LAEs at $z \sim 4.5$; their 
objects are either very young ($<15$ Myr) or very old ($>400$ Myr). Their results could be 
explained by invoking two different star formation modes in LAEs: a recent strong burst 
or a continuous SF in which the bulk of the population is dominated by old objects.
To better evaluate this possibility, we plot the normalized distribution 
of the mass-weighted stellar ages for all the LAEs identified in the simulation volume at 
$z\sim 5.7,6.6$ and $7.6$ in Fig. \ref{pdf}. We do not find any such bimodality from our simulation. 
LAEs are distributed in age between 39-165 Myr at $z\sim 7.6$ and this range increases with 
decreasing redshift, as already mentioned above.  A possible explanation of the experimental
result could be the presence of dust. LAEs would be visible in the Ly$\alpha$ when 
the age $<15$ Myr so that not enough dust would have formed and at later times $> 400$ Myr 
when the galaxy would have destroyed/ removed most of its dust content. In the intermediate 
periods, the Ly$\alpha$ line would be attenuated below observable limits. However, a complete 
picture is possible only if dust creation/destruction is modelled accurately, which we leave 
for future work.

Another interesting point is the Ly$\alpha$ line equivalent width. We calculate the rest frame 
observed EWs at $z \sim 5.7$ using our estimates of $f_c=0.22$ and $f_\alpha=0.3$ from the 
UV and Ly$\alpha$ LFs as mentioned in Tab. \ref{table1} as
$$EW = EW^{int} \frac{f_\alpha}{f_c}  (1-f_{esc}) {T_\alpha} ,$$
where $EW^{int} = L_\alpha^{int}/ L_c ^{int}(1375\, {\rm \AA})$. 
For our sample we find that $55 < ($EW/\AA$) < 89$, with mean value of 63.5\AA. This is
about a factor of 2 lower than the mean value of $120$ \AA, observed by Shimasaku et al. (2006). This discrepancy is probably due to physical effects that are not included in our model and which could broaden the EW distribution and increase its mean value. These include:
(i) Ly$\alpha$ emission from cooling of collisionally excited \HI in the ISM (Dijkstra 2009). 
Adding this contribution to the Ly$\alpha$ luminosity from stars could increase the intrinsic 
Ly$\alpha$ luminosity from the galaxy by large amounts while leaving the intrinsic continuum 
luminosity value unchanged, thereby increasing the intrinsic EW; (ii) we are using the same 
escape fractions of Ly$\alpha$ and continuum photons for all the galaxies. In reality, 
however, these values would change on a galaxy to galaxy basis, depending on the amount, 
topology and distribution of dust inside the galaxy, contributing to a broadening of the EW 
distribution; (iii) outflows could increase the observed EW distribution since they would 
enable more of the Ly$\alpha$ luminosity to escape undamped by dust, while leaving the 
continuum unchanged. Further, outflows with a sufficiently high value of the \HI column 
density ($\sim 10^{18-21} {\rm cm^{-2}} $) add a bump to the red part of the Ly$\alpha$ line, 
thereby increasing the observed EW (Verhamme et al. 2006); (iv) inflows would lead to a 
further damping of the Ly$\alpha$ luminosity and lead to a decrease in the EW. A full exploration of these effects requires modelling the amounts and distribution of dust and peculiar gas motions in individual galaxies. We defer this to future works at this point.

At $z \sim 5.7$, we find that for all galaxies with an observed Ly$\alpha$ luminosity, $L_\alpha \geq 10^{42.2} {\rm erg \, s^{-1}}$, the observed EW is greater than 20\AA. This means that all LBGs having a Ly$\alpha$ luminosity larger than this value show an observed Ly$\alpha$ EW $> 20$ \AA\, from our work. Although their samples are very small (6 and 8 galaxies respectively), Stanway et al. (2004) and Dow-Hygelund et al. (2007) find that only about 30\% of LBGs showing a Ly$\alpha$ emission line have an observed Ly$\alpha$ EW larger then 20 \AA. The main reason for this tension could be the fact that we are using the same ratio of $f_\alpha/f_c = 1.4$ for all the galaxies at $z \sim 5.7$; however, this value should change on a galaxy to galaxy basis. The relative escape fractions of Ly$\alpha$ and continuum photons must depend on the amount and distribution of dust and \HI inside the galaxy. Modelling this effect more realistically would probably lead to many galaxies having $EW<20$ \AA\, and resolve this tension.

LAEs are among the most distant galaxies known. As the quest for reionization sources 
is struggling to identify the most important populations for this process, it is
worthwhile to assess to what extent LAEs might represent such a long-searched-for
population. 
First we note that (Tab. \ref{table2}) at $z \sim 5.7$, the SFR density provided by LAEs
is $\dot\rho_* = 1.1 \times 10^{-2}\, {\rm M_\odot\, yr^{-1}\, Mpc^{-3}}$, and this decreases with increasing redshift. 
Comparing this to the cosmic SFR density of $0.12\, {\rm M_\odot\, yr^{-1}\, Mpc^{-3}}$ at the same redshift measured by Hopkins (2004), we find that LAEs contribute only about 9.2\% to the cosmic SFR value, confirming the previous 
result of DFG08. Corresponding to this SFR evolution and $f_{esc}=0.02$, it is easy to derive that the  
\HI ionizing photon rate density ($q_{LAE}$) contributed by LAEs is $q_{LAE}= 3.1\times 10^{49} \, {\rm s^{-1}\, 
Mpc^{-3}}$ at $z=6.6$. We compare this photon rate density to the one necessary to balance recombinations 
given by Madau, Haardt and Rees (1999):

\begin{equation}
q_{rec} = 10^{51.57} \bigg(\frac{C}{30}\bigg) \bigg(\frac{1+z}{7.6}\bigg)^3 
\bigg(\frac{\Omega_b h^2}{0.022}\bigg)^2 {{\rm s^{-1}\, Mpc^{-3}}},
\end{equation}
where $C$ is the IGM clumping factor.  For $C=1$, corresponding to a homogeneous IGM, we get 
the minimum \HI ionizing photon rate density necessary to balance recombinations,  i.e. 
$q_{rec} = 1.24 \times 10^{50} \, {\rm s^{-1}\, Mpc^{-3}}$. This means that the LAEs on 
the Ly$\alpha$ LF at $z\sim 6.6$ can contribute at most 25\% of the \HI ionizing photons 
needed to balance recombinations at this redshift. This value decreases to 0.8\% as $C$ increases 
to 30. 

The total \HI ionizing photon rate density contributed by all astrophysical sources ($q_{all}$) 
at $z \sim 6.6$ can be calculated as (Bolton \& Haehnelt, 2007)
\begin{equation}
q_{all} = 10^{51.13} \Gamma_{-12} \bigg(\frac{1+z}{7.6}\bigg)^{-2} 
{{\rm s^{-1}\, Mpc^{-3}}};
\end{equation}
using $\Gamma_{-12} = 0.19$ (Tab. \ref{table1}), gives $q_{all}= 2.5 \times 10^{50}\, {\rm s^{-1}\, Mpc^{-3}}$ 
at $z \sim 6.6$.  This implies LAEs contribute about 12.5\% of the total \HI ionizing photon rate, which is consistent with the SFR density contribution of LAEs to the global value. Such a low contribution from LAE, given their halo and stellar masses does not come as a surprise. Indeed, this result is consistent with the previous estimates by Choudhury \& Ferrara (2007) who showed that only a fraction $\simlt 1$\% of the photons required to ionize the IGM come from objects in the LAE range, the bulk being provided at high redshifts by faint (or even Ultra Faint, see Salvadori \& Ferrara 2009) dwarf galaxies.    

\section{Summary}
\label{Concl}
We use a large scale hydrodynamical simulation to derive the physical properties 
(halo, stellar, gas mass, star formation rate, stellar age and metallicity) of 
high redshift galaxies. Using a reionization history consistent with an early 
reionization epoch (ERM), we fix the evolution of the background photoionization 
rate, and calculate the size of the Str\"omgren sphere around each object, taking 
into account the ionization boost due to spatial clustering of the sources when 
computing the corresponding Ly$\alpha$ radiation transmission. 
We then define as LAEs galaxies with an observed Ly$\alpha$ luminosity above the 
currently observable limit ($L_\alpha \geq 10^{42.2} {\rm erg \, s^{-1}}$) and an observed EW $>20$ \AA. 

We find that Ly$\alpha$ photons at both $z \sim 5.7$ and $6.6$ must be attenuated by 
dust within the galaxy to reproduce the observed LAE luminosity function. The amount
of photons that escape the ISM, undamped by dust is $f_\alpha=0.3$ at both redshifts. At $z \sim 6.6$ the simulated
LF lies between the upper and lower limits of the data. At face value, this results 
implies that further spectroscopy of LAE candidates at this redshift should rule out 
$\sim 60$\% of them as LAEs. 

Clustering of sources boosts the average value of the total photoionization rate (and 
consequently the ionized hydrogen fraction) in the surroundings of galaxies by $<1.5$ 
times at $z \sim 5.7,6.6$; this value increases by more than a factor of 50 at 
$z \sim 7.6$. For an almost neutral IGM ($z \sim 7.6, \chi_{HI}=0.16$), if clustering effects are ignored, only those few galaxies that are able to carve out a large enough \HII region are visible. However, about 4 times more galaxies become visible when clustering effects are included. In a highly ionized IGM ($z\sim 6.6, 5.7$, $\chi_{HI} \sim 10^{-4}, 10^{-5}$), however, the 
effect of clustering is indiscernible on the Ly$\alpha$ LF. Note that, the clustering boost 
is not necessarily the highest for the most massive galaxies. Our model is probably 
somewhat overstating the effects of clustering because we are not treating the 
radiation transfer of ionizing radiation.


The fraction of Ly$\alpha$ luminosity transmitted always increases with SFR and about 
$50$\% of the Ly$\alpha$ luminosity is transmitted at $z=5.6,6.6$; however, this 
relation shows a larger scatter at $z=7.6$ (transmission between 20-50 \%) compared to 
the lower redshifts since at $z \sim 7.6$, it is essentially the photoionization rate 
boost due to clustering which enables Ly$\alpha$ luminosity transmission, if galaxies 
are embedded in the neutral patches of the IGM.

The presence of dust even at these high redshift is further confirmed by the analysis
of the LAE UV luminosity function, which we have also reproduced within our study. 
Matching the UV LF requires that the escape fraction of UV continuum photons is 
$f_c=(0.22,0.37)$ at $z =(5.7,6.6)$, again a clear signature of dust attenuation. 
We interpret the higher attenuation of the continuum photons relative to the Ly$\alpha$ 
(Neufeld 1991) at $z \sim 5.7$ as hinting at an inhomogeneous two-phase ISM with dust clumps embedded 
in a warm intercloud gas while the relative attenuation is consistent with homogeneously distributed dust 
at $z \sim 6.6$. At $z \sim 5.7$, the color excess calculated from our model, $E(B-V) = 0.15$, 
is consistent with the value deduced by Lai et al. (2007), $E(B-V) < 0.425$, 
and the results obtained by Nagamine et al. (2007) ($E(B-V) = 0.15$) 
using SPH simulations. The color excess, $E(B-V)=0.1$, at $z \sim 6.6$ is 
lower than the value at $z\sim 5.7$; this is expected due to to the average lower 
metallicities of LAEs towards higher redshifts (see Tab. 3). By extrapolating these 
results at $z=7.6$, we have predicted the UV and Ly$\alpha$ LFs for LAEs at $z \sim 7.6$. 

As a further test of our results, we have selected three LAEs from our simulation whose 
observed Ly$\alpha$ luminosities match most closely with the values observed at $z \sim 5.7$ 
by Lai et al. (2007). For each of the chosen objects, the intrinsic spectra (obtained using 
{\rm STARBURST99}) is attenuated using a SN dust extinction curve and $E(B-V)=0.15$ to match 
to the observations. We find our synthetic Spectral Energy Distributions are in very good 
agreement with the observations. The objects observed are found to have ages of a few hundred 
Myr, metallicity about $20-30$\% $Z_\odot$ and SFR of 7-10 ${\rm M_\odot}\, {\rm yr^{-1}}$. All three 
LAEs are well fit by a single value of the color excess, hinting at the fact they all might 
be dust enriched to a similar degree.

Finally, we have discussed in detail the physical properties (and we provide handy fitting 
functions to several relations among them) of galaxies which we identify 
as LAEs at $z=5.6, 6.6$ and $7.6$. The ages of the LAEs 
range between $39-165$ Myr at $z \sim 7.6$ but this range increases to $38-326$ Myr at 
$z\sim 5.7$. Further, the average metallicity of LAEs is $Z=0.12 Z_\odot$ at $z\sim 7.6$, 
and increases to $Z=0.22 Z_\odot$ at $z\sim 5.7$. Hence, LAEs are more metal-enriched than 
what usually assumed. Star formation is relatively suppressed by feedback in low mass halos 
($M_h \leq 10^{11} M_\odot$) and it rises steeply for the larger halo masses. 
The large majority of LAEs seem to have had a higher average star formation rate over the entire history as compared to the final value. Further study would be necessary to put this statement on more solid grounds.


\section*{Acknowledgments} 
We deeply appreciate the thoroughness and the insight of the referee. We thank N. Kashikawa, M. Kobayashi, S. Salvadori for helpful discussions and insightful 
comments, and S. Bianchi for providing SN dust extinction curves. It is a pleasure to 
acknowledge (PD and AF) the warm hospitality by NAOJ, Mitaka where part of this work has 
been carried on. The simulations presented here have been carried out at the CINECA 
Supercomputing Center, with CPU time allocated within the INAF-CINECA agreement.  
AS, SB and LT acknowledge financial support from the PRIN-2007 MIUR Grant "The Cosmic 
Cycle of Baryons".



\newpage 
\label{lastpage} 
\end{document}